\documentclass[review,12pt]{elsarticle}

\usepackage{amssymb}
\usepackage{booktabs}
\usepackage{amsmath,amsfonts}
\usepackage{times}
\usepackage{lineno}

\journal{arXiv}

\begin{document}

\begin{frontmatter}



\title{Scattering-Matrix-Based Parametric Characterization of a Two-Port Bridged-T Network for Microstrip Filter Applications}


\author[add1]{Naser Khatti Dizabadi \corref{cor1}}\ead{nak5300@utulsa.edu}\cortext[cor1]{Corresponding author}
\author[add1]{Douglas Jussaume}\ead{douglas-jussaume@utulsa.edu} 
%\author[add1]{Kaveh Ashenayi}\ead{kash@utulsa.edu}
%\author[add1]{Heng-Ming Tai}\ead{tai@utulsa.edu}

\affiliation[add1]{organization={Department of Electrical and Computer Engineering, The University of Tulsa},%Department and Organization
            addressline={800 S. Tucker Drive}, 
            city={Tulsa},
            postcode={74104}, 
            state={OK},
            country={USA}}

\begin{abstract}
The purpose of this study is to characterize a two-port Bridged-T network using transmission (T) and scattering (S) matrices. Using mathematical derivations, scattering parameters including S11, S12, S21, and S22 have been derived from the T and S matrices to permit a detailed investigation of the network's performance. As two of the most relevant parameters in the design of microstrip filters, both the magnitude and phase of S11 and S21 have been parametrically calculated after normalizing the frequency. Furthermore, when the inductors L1 and L2 are identical, all even coefficients of the numerator polynomial in the S11 transfer function are eliminated, leaving only the odd coefficients behind. Based on this feature, the bridged-T circuit is designed to operate as a high-pass filter. Therefore, the magnitude and phase of both S11 and S21 have been simulated for the designed filter with a corner frequency of 1 GHz. Simulation results performed by Keysight ADS show that S11 and S21 for the high-pass filter built upon the bridged-T network have sharp roll-off ratios of -30dB/GHz and -32dB/GHz respectively.
\end{abstract}

%\includegraphics[width=1\linewidth]{img/myImage.PDF}



\end{frontmatter}

%\linenumbers

\section{Introduction}
\label{intr}
High-frequency microstrip filters are widely used in the design and construction of radio communication systems \cite{Trans-1, Trans-2, Trans-3, Trans-4, Trans-5, Trans-6, MyVCO1, MyVCO2}. The versatility of microstrip filters is derived from their ability to combine a variety of subsystems with unique transfer functions. This allows for the manipulation of the transfer function by strategically varying the geometric locations of the poles and zeros in a complex frequency plane. A bridged-T two-port network is one of the most commonly used subsystems in both low-frequency and high-frequency circuits.

While several studies have examined the use of different bridged-T networks in the design of microstrip filters \cite{BT-1, BT-2, BT-3, BT-4, BT-5, BT-6, BT-7}, this paper explores the characterization of a bridged-T network enhanced with a parallel LC resonator. This study aims to extract all components of the transmission and scattering matrices in order to provide a more comprehensive understanding of its behavior under a variety of circumstances.

As a further step, we parametrically determine both the magnitude and phase of the critical scattering parameters, S11 and S21. These parameters are instrumental in evaluating the quality of a microstrip filter. By introducing two identical inductors into the bridge section, we demonstrate that the bridged-T network can be transformed into an effective high-pass filter.

In order to verify this finding, we meticulously designed and simulated a high-pass filter with a cutoff frequency of 1 GHz. A sharp roll-off is observed in the simulation results, confirming the bridged-T network's ability to act as a high-pass filter.

Detailed explanations and simulation results follow, providing a comprehensive overview of the bridged-T network's performance and its significance for high-pass filtering.

\section{The Transmission and Scattering Matrices}
\label{S-Matrix}

\begin{figure}[]
\centering
\noindent
\includegraphics[width=3.5in]{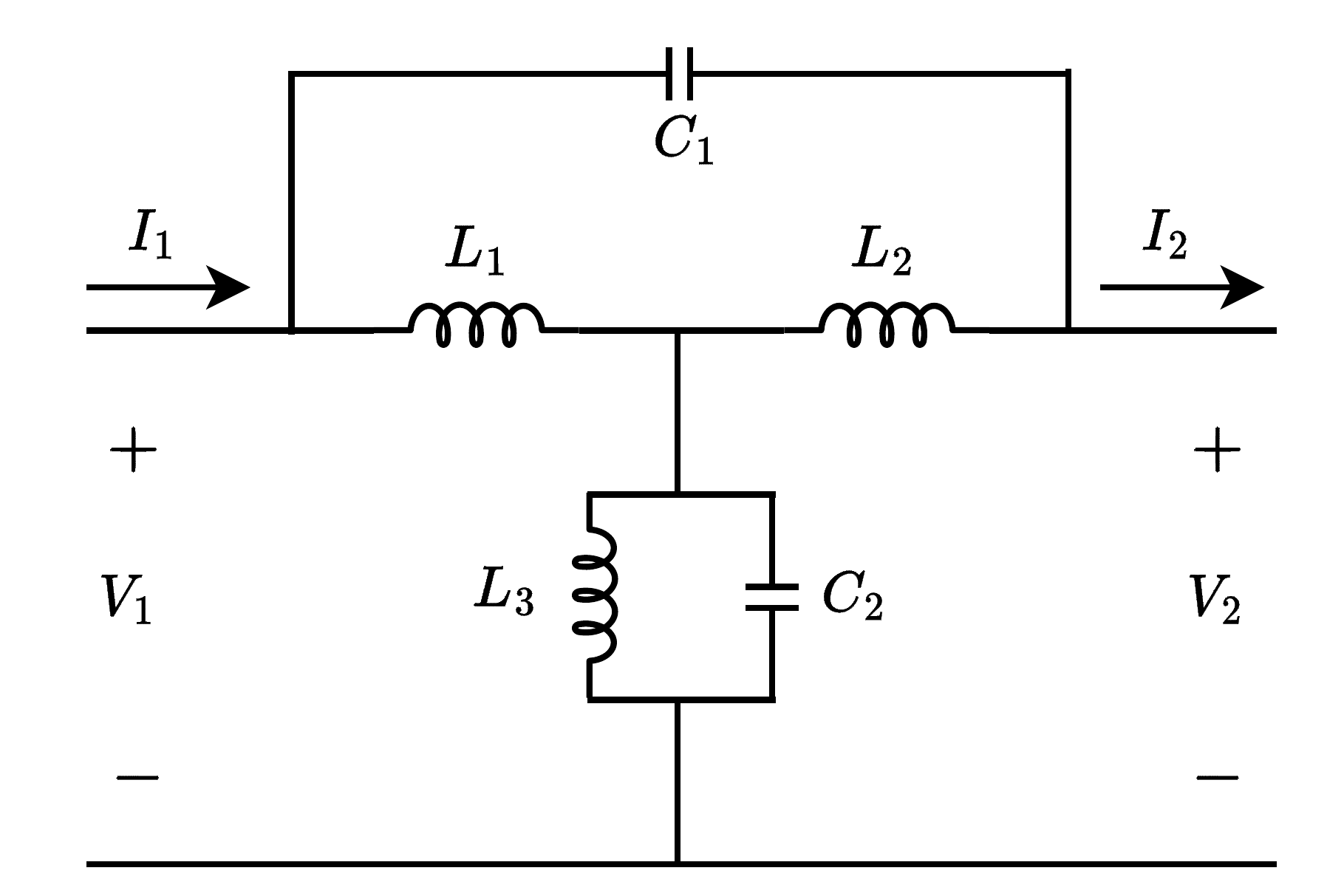}
\caption{A two-port capacitive bridged-T network}
\label{fig-1}
\end{figure}

Figure \ref{fig-1} shows a customary two-port Bridged-T network in which capacitor \(C_{1}\) establishes a bridge connection between port one and port two, while the combination of the inductors \(L_{1}\) and \(L_{2}\), along with the parallel LC tank, create a T sub-network. The parallel LC tank, comprised of capacitor \(C_{2}\) and inductor \(L_{3}\), assumes the role of determining the resonance frequency within this circuit.

To begin the analysis of the circuit depicted in Figure \ref{fig-1}, the Laplace impedances of all components have been calculated and are presented in Equations (\ref{Z_{L1}}) to (\ref{Z_{L3C2}}). 

\begin{equation}
\label{Z_{L1}}
Z_{L_{1}} = SL_{1}
\end{equation}

\begin{equation}
\label{Z_{L2}}
Z_{L_{2}} = SL_{2}
\end{equation}

\begin{equation}
\label{Z_{L3}}
Z_{L_{3}} = SL_{3}
\end{equation}

\begin{equation}
\label{Z_{L3C2}}
Z_{L_{3}C_{2}} = \frac{1}{\frac{1}{SL_{3}}+SC_{2}}
\end{equation}

\begin{figure}[]
\centering
\noindent
\includegraphics[width=3.5in]{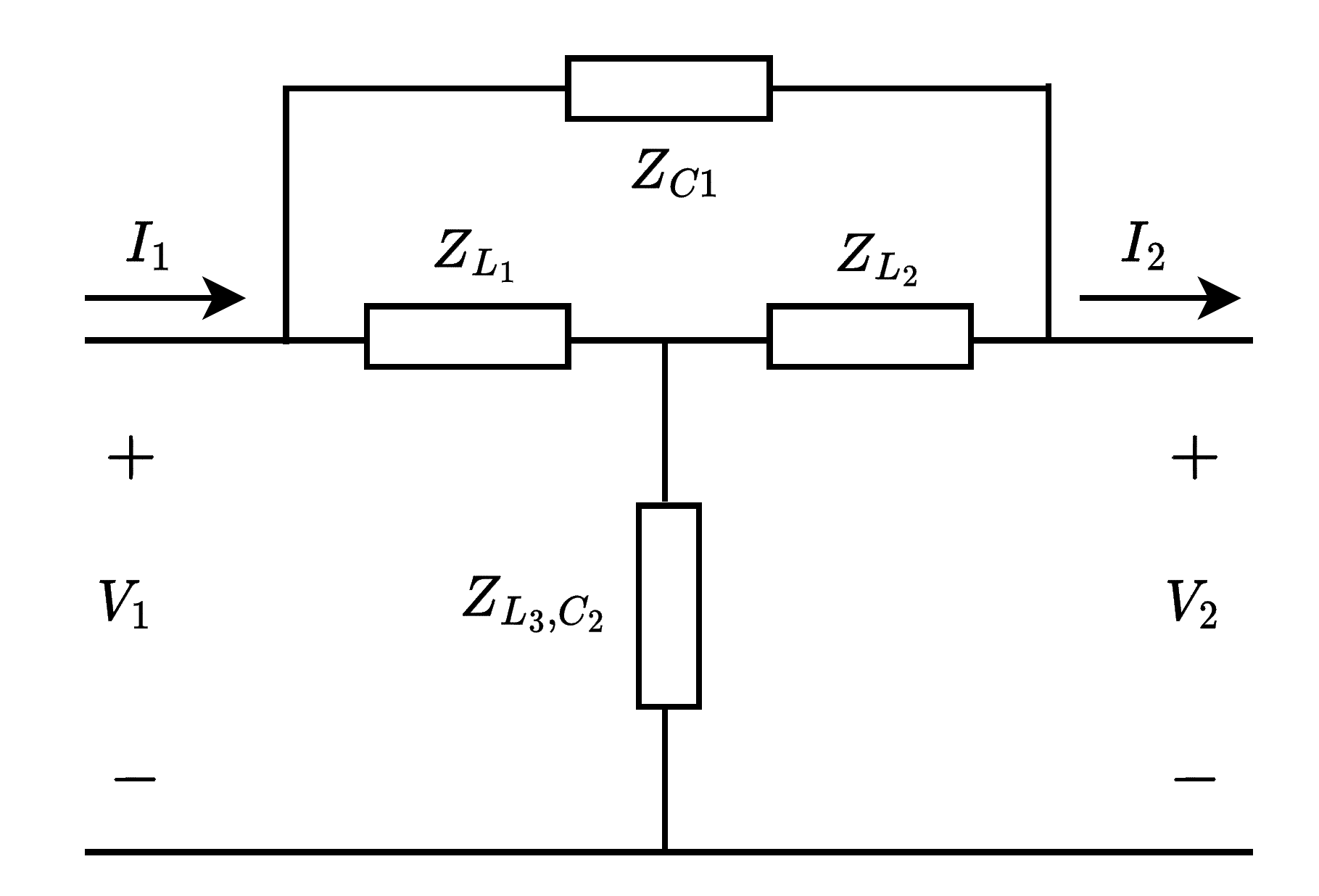}
\caption{Laplace equivalent impedance of the bridged-T network}
\label{fig-2}
\end{figure}

To see better, an illustration of the equivalent impedances of the components is provided in Figure \ref{fig-2}.

\begin{figure}[!t]
\centering
\noindent
\includegraphics[width=3.5in]{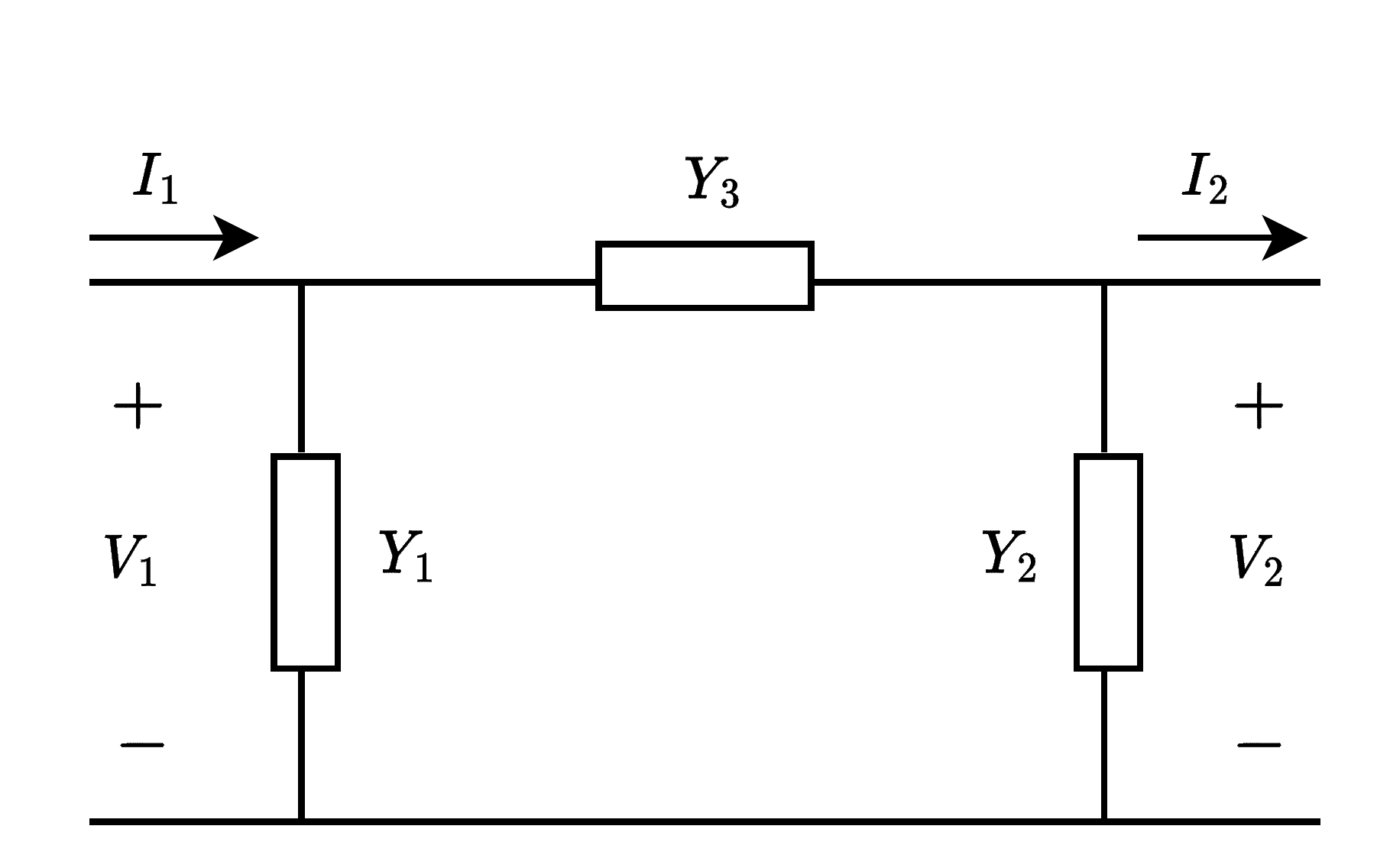}
\caption{Laplace equivalent impedance of the bridged-T network}
\label{fig-3}
\end{figure}

After applying a Y to \(\Delta\) conversion, as described in \cite{Y-Delta}, to the circuit depicted in Figure \ref{fig-2}, the bridged-T network has been transformed into a \(\Pi\) network, simplifying the analysis. This simplified \(\Pi\) network is illustrated in Figure \ref{fig-3} where \(Y_{1}\), \(Y_{2}\), and \(Y_{3}\) are the equivalent admittances for each component in the \(\Pi\) network equated through the equations (\ref{Y_{1}}) to (\ref{Y_{3}}).

\begin{equation}
\label{Y_{1}}
Y_{1} =\frac{1}{ SL_{1}+\frac{L_{3}(1+\frac{L_{1}}{L_{2}})}{1+S^{2}C_{2}L_{3}}}
\end{equation}

\begin{equation}
\label{Y_{2}}
Y_{2} =\frac{1}{ SL_{2}+\frac{L_{3}(1+\frac{L_{2}}{L_{1}})}{1+S^{2}C_{2}L_{3}}}
\end{equation}

\begin{equation}
\label{Y_{3}}
Y_{3} =\frac{1}{\frac{1}{SC_{1}}+S(L_{1}+L_{2})+L_{1}L_{2}(\frac{S}{L_{3}}+S^3C_{2})}
\end{equation}\\

According to \cite{Pozar-P190}, a transmission (or ABCD or T for short) matrix as equated in equation (\ref{T-1}) can be directly derived from the admittances in a \(\Pi\) network. In this equation, A, B, C, and, D are the T matrix's components and can be calculated through the converting equations (\ref{A}) to (\ref{D}).

%\vspace{10pt} % Add 10 points of vertical space
\begin{equation}
\label{T-1}
\begin{bmatrix}
V_1 \\
I_1
\end{bmatrix}
=
\begin{bmatrix}
A & B \\
C & D
\end{bmatrix}
\begin{bmatrix}
V_2 \\
I_2 \\
\end{bmatrix}
\end{equation}
%\vspace{10pt} % Add 10 points of vertical space

\begin{equation}
\label{A}
A =1 + \frac{Y_{2}}{Y_{3}}
\end{equation}

\begin{equation}
\label{B}
B = \frac{1}{Y_{3}}
\end{equation}

\begin{equation}
\label{C}
C = Y_{1} + Y_{2} + \frac{Y_{1}Y_{2}}{Y_{3}}
\end{equation}

\begin{equation}
\label{D}
D = 1 + \frac{Y_{1}}{Y_{3}}
\end{equation}
%\vspace{10pt} % Add 10 points of vertical space

In accordance with \cite{Pozar-P192}, the conversion of the T matrix to the S matrix can be achieved using the following set of conversion formulas equated by equations (\ref{S11}) to (\ref{S22}). In these equations, \(Z_{0}\) represents the characteristic impedance of a typical transmission line.

%\vspace{10pt} % Add 10 points of vertical space
\begin{equation}
\label{S11}
\begin{aligned}
S_{11} &= \frac{A + B / Z_{o} - CZ_{o} - D}{A + B/ Z_{o} + CZ_{o} + D} 
\end{aligned}
\end{equation}

\begin{equation}
\label{S12}
\begin{aligned}
S_{12} &=  \frac{2(AD-BC)}{A + B/ Z_{o} + CZ_{o} + D}
\end{aligned}
\end{equation}

\begin{equation}
\label{S21}
\begin{aligned}
S_{21} &=  \frac{2}{A + B/ Z_{o} + CZ_{o} + D}
\end{aligned}
\end{equation}

\begin{equation}
\label{S22}
\begin{aligned}
S_{22} &=  \frac{-A + B / Z_{o} - CZ_{o} + D}{A + B/ Z_{o} + CZ_{o} + D}
\end{aligned}
\end{equation}
%\vspace{10pt} % Add 10 points of vertical space

In the process of microstrip filter design, the parameters S21 and S11 assume considerable significance. Accordingly, we embark on the computation of the canonical forms of their respective transfer functions. To this end, the substitution of equations (\ref{A}) through (\ref{D}) into equations (\ref{S11}) through (\ref{S22}), followed by necessary simplifications, results in the formulation of a simplified version of the S11 transfer function, denoted as equation \ref{S11-1}.

%\vspace{10pt} % Add 10 points of vertical space
\begin{equation}
\label{S11-1}
\begin{aligned}
S_{11} &= \frac{N_{0}+N_{1}S+N_{2}S^2+N_{3}S^3+N_{4}S^4}{D_{0}+D_{1}S+D_{2}S^2+D_{3}S^3+D_{4}S^4+D_{5}S^5}
\end{aligned}
\end{equation}
%\vspace{10pt} % Add 10 points of vertical space

where coefficients \(N_{0}\), \(N_{1}\), \(N_{2}\), \(N_{3}\), \(N_{4}\), \(D_{0}\), \(D_{1}\), \(D_{2}\), \(D_{3}\), \(D_{4}\), and \(D_{5}\) are as follows:

%\vspace{10pt} % Add 10 points of vertical space
\begin{equation}
\label{S11-2}
\begin{aligned}
N_{0} &= C_{1}L_{3}Z_{0}^2 \\ 
N_{1} &= C_{1}L_{3}Z_{0}(L_{1} -L_{2}) \\ 
%N_{2} &= C_{1}(L_{1} + L_{2} -C_{2}Z_{0}^2) + L_{3}(-L_{1}Z_{0}^2 - L_{2}Z_{0}^2 + C_{1}L_{1}L_{2})  \\
N_{2} &= (C_{1}-L_{3}Z_{0}^2)(L_{1} + L_{2}) + C_{1}(L_{1}L_{2}L_{3} -C_{2}Z_{0}^2)  \\
%N_{3} &= C_{1}C_{2}L_{1}Z_{0} - C_{1}C_{2}L_{2}Z_{0} \\ 
N_{3} &= C_{1}C_{2}Z_{0}(L_{1} - L_{2}) \\
%N_{4} &= C_{1}C_{2}L_{1}L_{2} - C_{2}L_{2}Z_{0}^2 - C_{2}L_{1}Z_{0}^2 \\
N_{4} &= C_{2}(C_{1}L_{1}L_{2} -Z_{0}^2(L_{1} + L_{2})) \\
D_{0} &= C_{1}L_{3}Z_{0}^2 \\
%D_{1} &= 2C_{1}Z_{0} + C_{1}L_{1}L_{3}Z_{0} + C_{1}L_{2}L_{3}Z_{0} \\
D_{1} &= C_{1}Z_{0}(2 + L_{3}(L_{1} + L_{2})) \\
%D_{2} &= C_{1}L_{1} + C_{1}L_{2} + C_{1}C_{2}Z_{0}^2 + L_{1}L_{3}Z_{0}^2 + L_{2}L_{3}Z_{0}^2 + C_{1}L_{1}L_{2}L_{3} \\
D_{2} &= (C_{1}+L_{3}Z_{0}^2)(L_{1}+L_{2})+C_{1}(L_{1}L_{2}L_{3} + C_{2}Z_{0}^2) \\
%D_{3} &= 2L_{1}Z_{0} + 2L_{2}Z_{0} + C_{1}C_{2}L_{1}Z_{0} + C_{1}C_{2}L_{2}Z_{0} + 2L_{1}L_{2}L_{3}Z_{0} \\
D_{3} &= ((2+C_{1}C_{2})(L_{1}+L_{2})+2L_{1}L_{2}L_{3})Z_{0} \\
%D_{4} &= C_{2}L_{1}Z_{0}^2 + C_{2}L_{2}Z_{0}^2 + C_{1}C_{2}L_{1}L_{2} \\
D_{4} &= C_{2}(C_{1}L_{1}L_{2} + Z_{0}^2(L_{1} + L_{2})) \\
D_{5} &= 2L_{1}L_{2}C_{2}Z_{0} \\
\end{aligned}
\end{equation}
%\vspace{10pt} % Add 10 points of vertical space

Similarly, the simplified version of the S21 transfer function is expressed by equation \ref{S21-1}.

%\vspace{10pt} % Add 10 points of vertical space
\begin{equation}
\label{S21-1}
\begin{aligned}
S_{21} &= \frac{N'_{1}S+N'_{3}S^3+N'_{5}S^5}{D_{0}+D_{1}S+D_{2}S^2+D_{3}S^3+D_{4}S^4+D_{5}S^5}
\end{aligned}
\end{equation}
%\vspace{10pt} % Add 10 points of vertical space

where coefficients \(N'_{1}\), \(N'_{3}\), and \(N'_{5}\) are:

%\vspace{10pt} % Add 10 points of vertical space
\begin{equation}
\label{S21-2}
\begin{aligned}
N'_{1} &= 2C_{1}Z_{0} \\ 
N'_{3} &= 2Z_{0}(L_{1} + L_{2} + L_{1}L_{2}L_{3}) \\ 
N'_{5} &= 2C_{2}L_{1}L_{2}Z_{0} \\ 
%D_{0} &= C_{1}L_{3}Z_{0}^2 \\
%D_{1} &= 2C_{1}Z_{0} + C_{1}L_{1}L_{3}Z_{0} + C_{1}L_{2}L_{3}Z_{0} \\
%D_{2} &= C_{1}L_{1} + C_{1}L_{2} + C_{1}C_{2}Z_{0}^2 + L_{1}L_{3}Z_{0}^2 + L_{2}L_{3}Z_{0}^2 + C_{1}L_{1}L_{2}L_{3} \\
%D_{3} &= 2L_{1}Z_{0} + 2L_{2}Z_{0} + C_{1}C_{2}L_{1}Z_{0} + C_{1}C_{2}L_{2}Z_{0} + 2L_{1}L_{2}L_{3}Z_{0} \\
%D_{4} &= C_{2}L_{1}Z_{0}^2 + C_{2}L_{2}Z_{0}^2 + C_{1}C_{2}L_{1}L_{2}\\
%D_{5} &= 2C_{2}L_{1}L_{2}Z_{0} \\
\end{aligned}
\end{equation}
%\vspace{10pt} % Add 10 points of vertical space

Let's say that the complex frequency, denoted as \(S\), is equal to \(j\omega\), where \(\omega\) is the angular frequency. Additionally, in the design of a filter, the normalized angular frequency, denoted as \(\Omega\), is defined as the ratio of the angular frequency (\(\omega\)) to the cut-off angular frequency (\(\omega_{c}\)). With these definitions in place, the transfer functions of S11 and S21, as given by equations (\ref{S11-1}) and (\ref{S21-1}), can be reformulated in a normalized and complex form as follows:

%\vspace{10pt} % Add 10 points of vertical space
\begin{equation}
\label{S11-simplified-1}
\begin{aligned}
S_{11} &= \frac{\alpha_{N} + j\beta_{N}}{\alpha_{D} + j\beta_{D}}
\end{aligned}
\end{equation}
%\vspace{10pt} % Add 10 points of vertical space

where the coefficients \(\alpha_{N}\), \(\beta_{N}\), \(\alpha_{D}\), and \(\beta_{D}\) are obtained through the following parametric expressions: 

%\vspace{10pt} % Add 10 points of vertical space
\begin{equation}
\label{S11-simplified-2}
\begin{aligned}
\alpha_{N} &= N_{0} - N_{2}\omega_{c}^2\Omega^2 + N_{4}\omega_{c}^4\Omega^4\\
\beta_{N} &= N_{1}\omega_{c}\Omega - N_{3}\omega_{c}^3\Omega^3\\
\alpha_{D} &= D_{0} - D_{2}\omega_{c}^2\Omega^2 + D_{4}\omega_{c}^4\Omega^4\\
\beta_{D} &= D_{1}\omega_{c}\Omega - D_{3}\omega_{c}^3\Omega^3 +  D_{5}\omega_{c}^5\Omega^5
\end{aligned}
\end{equation}
%\vspace{10pt} % Add 10 points of vertical space

By using Equation (\ref{S11-simplified-1}), both the magnitude and phase of S11 can be calculated as follows:

%\vspace{10pt} % Add 10 points of vertical space
\begin{equation}
\label{S11-simplified-3}
\begin{aligned}
\left| S_{11} \right| &= \sqrt{\frac{(\alpha_{N})^2 + (\beta_{N})^2}{(\alpha_{D})^2 + (\beta_{D})^2}}
\end{aligned}
\end{equation}
%\vspace{10pt} % Add 10 points of vertical space

%\vspace{10pt} % Add 10 points of vertical space
\begin{equation}
\label{S11-simplified-4}
\begin{aligned}
\angle S_{11} &= \tan^{-1}(\frac{ \beta_{N}}{\alpha_{N}}) - \tan^{-1}(\frac{\beta_{D}}{\alpha_{D}})
\end{aligned}
\end{equation}
%\vspace{10pt} % Add 10 points of vertical space

In a similar manner, after normalizing the frequency and applying some simplifications, the complex form of the S21 transfer function can also be obtained as shown in Equation (\ref{S21-simplified-1}).

%\vspace{10pt} % Add 10 points of vertical space
\begin{equation}
\label{S21-simplified-1}
\begin{aligned}
S_{21} &= \frac{\alpha'_{N} + j\beta'_{N}}{\alpha_{D} + j\beta_{D}}
\end{aligned}
\end{equation}
%\vspace{10pt} % Add 10 points of vertical space

where the \(\alpha'_{N}\) and \(\beta'_{N}\) denote the respective parametric real and imaginary components associated with the numerator in the complex representation of the S21 transfer function, as articulated below:

%\vspace{10pt} % Add 10 points of vertical space
\begin{equation}
\label{S21-simplified-2}
\begin{aligned}
\alpha'_{N} &= 0\\
\beta'_{N} &= N'_{1}\omega_{c}\Omega - N'_{3}\omega_{c}^3\Omega^3 + N'_{5}\omega_{c}^5\Omega^5\\
%\alpha'_{D} &= D'_{0} - D'_{2}\omega_{c}^2\Omega^2 + D'_{4}\omega_{c}^4\Omega^4\\
%\beta'_{D} &= D'_{1}\omega_{c}\Omega - D'_{3}\omega_{c}^3\Omega^3 +  D'_{5}\omega_{c}^5\Omega^5
\end{aligned}
\end{equation}
%\vspace{10pt} % Add 10 points of vertical space

Likewise, the magnitude and phase of the S21 in equation (\ref{S21-simplified-1}) can be determined through the following calculations:

%\vspace{10pt} % Add 10 points of vertical space
\begin{equation}
\label{S21-simplified-3}
\begin{aligned}
\left| S_{21} \right| &= \frac{\left| \beta'_{N} \right|}{\sqrt{(\alpha_{D})^2 + (\beta_{D})^2}}
\end{aligned}
\end{equation}
%\vspace{10pt} % Add 10 points of vertical space

%\vspace{10pt} % Add 10 points of vertical space
\begin{equation}
\label{S21-simplified-4}
\begin{aligned}
\angle S_{21} &= \pm \frac{\pi}{2} - \tan^{-1}(\frac{\beta_{D}}{\alpha_{D}})
\end{aligned}
\end{equation}
%\vspace{10pt} % Add 10 points of vertical space

\section{Design and Simulation of A High-Pass Filter}
\label{Design}

\begin{figure}[]
\centering
\noindent
\includegraphics[width=3.5in]{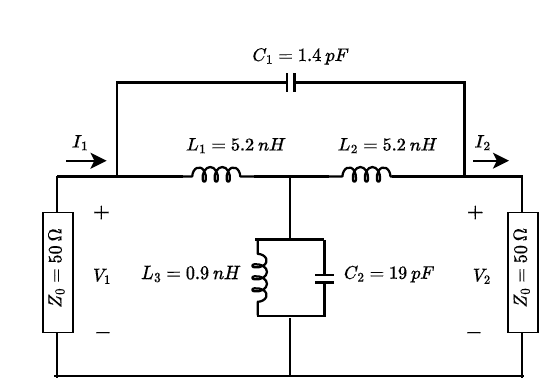}
\caption{High pass filter performance using a bridged-T network}
\label{fig-4}
\end{figure}

In this section, a high-pass filter was designed based on the bridged-T circuit illustrated in Figure \ref{fig-1}. In the design process, the inductors L1 and L2 were deliberately chosen to be identical, resulting in the elimination of all odd coefficients in the numerator of Equation (\ref{S11-1}), leaving only the even coefficients. By selecting the component values as shown in Figure \ref{fig-4}, the coefficients of the transfer functions for S11 and S21 were subsequently updated as shown in Equation (\ref{S11-simplified-numerical-2}). 

% Numerator of S11:  1 + 7428.58 S^2 + 156.825 S^4
% Denuminator of S11: 1 + 0.252444 S + 7428.61 S^2 + 1102.73 S^3 + 156.826 S^4 + 16.3098 S^5

% Numerator of S21: 0.0444444 S + 1102.73 S^3 + 16.3098 S^5
% Denuminator of S21: 1 + 0.252444 S + 7428.61 S^2 + 1102.73 S^3 + 156.826 S^4 + 16.3098 S^5

\vspace{10pt} % Add 10 points of vertical space
\begin{equation}
\label{S11-simplified-numerical-2}
\begin{aligned}
\alpha_{N} &\approx 1 - 7428.58\omega_{c}^2\Omega^2 + 156.825\omega_{c}^4\Omega^4\\
\beta_{N} &= 0\\
\alpha'_{N} &= 0\\
\beta'_{N} &\approx 0.044\omega_{c}\Omega - 1102.73\omega_{c}^3\Omega^3 + 16.31\omega_{c}^5\Omega^5\\
\alpha_{D} &\approx 1 - 7428.61\omega_{c}^2\Omega^2 + 156.826\omega_{c}^4\Omega^4\\
\beta_{D} &\approx 0.252\omega_{c}\Omega - 1102.73\omega_{c}^3\Omega^3 + 16.31\omega_{c}^5\Omega^5
\end{aligned}
\end{equation}
\vspace{10pt} % Add 10 points of vertical space

where \(\omega_{c}\) is the corner frequency and \(\Omega\) is the unitless normalized frequency for the designed high-pass filter.

\begin{figure}[]
\centering
\noindent
\includegraphics[width=3.5in]{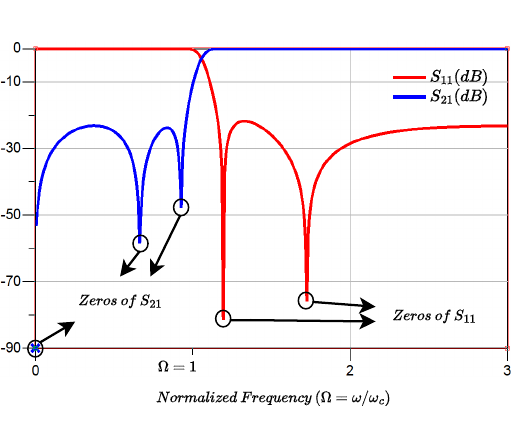}
\caption{\(S_{11}\) and \(S_{21}\) versus normalized frequency}
\label{fig-5}
\end{figure}

After establishing the corner frequency at about 1 GHz, Figure \ref{fig-5} illustrates the simulated S11 and S21 responses of the designed high-pass filter in relation to the normalized frequency. The results depicted in Figure \ref{fig-5} demonstrate that the bridged-T configuration, with symmetrically equal inductances L1 and L2, functions effectively as a high-pass filter, exhibiting a discernible sharp roll-off at the specified corner frequency.

\begin{figure}[]
\centering
\noindent
\includegraphics[width=3.5in]{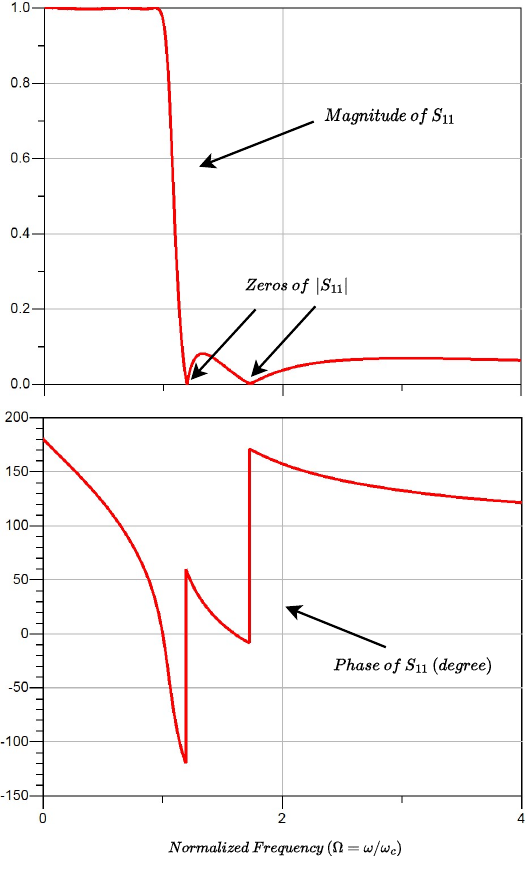}
\caption{Magnitude and phase of \(S_{11}\) versus the normalized frequency}
\label{fig-6}
\end{figure}

\begin{figure}[]
\centering
\noindent
\includegraphics[width=3.5in]{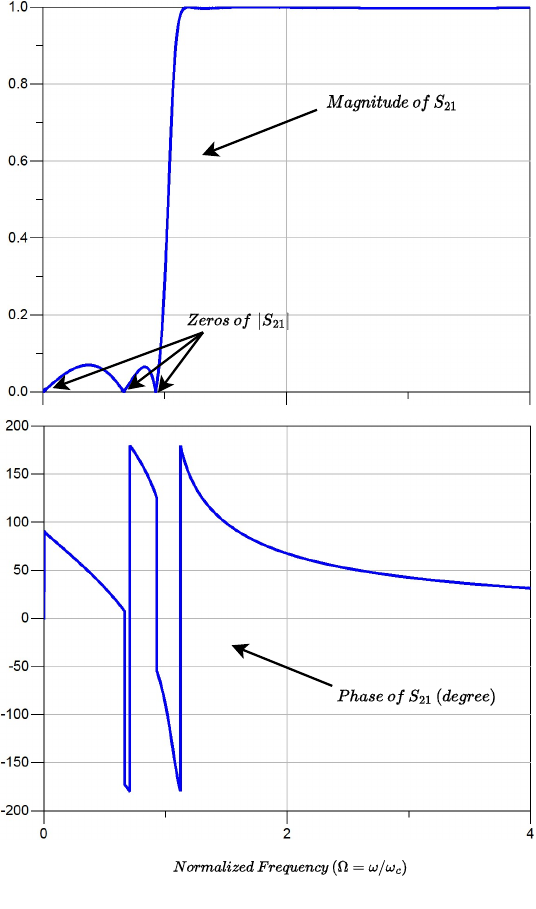}
\caption{Magnitude and phase of \(S_{21}\) versus normalized frequency}
\label{fig-7}
\end{figure}

To assess the congruence between the filter's behavior and mathematical predictions for both the magnitudes and phases of S11 and S21, Figures \ref{fig-6} and \ref{fig-7} illustrate these parameters with normalized frequency.

%Additionally, Figure \ref{fig-8} provides the simulated group delay for the designed high-pass filter based on the bridged-T two-port network.

%\begin{figure}[!t]
%\centering
%\includegraphics{Fig-8-v2-Normalized.pdf}
%\caption{Group Delay versus normalized frequency}
%\label{fig-8}
%\end{figure}

\section{Conclusion}
\label{Con}
This study characterizes a two-port Bridged-T network using transmission (T) and scattering (S) matrix techniques. The derived scattering parameters (S11, S12, S21, and S22) enable a detailed exploration of the bridged-T network's performance. Emphasizing the relevance of S11 and S21 in microstrip filter design, their magnitudes and phases are parametrically calculated after frequency normalization. Using this mathematical derivation, designers can alter the locations of the zeros for both S11 and S21 to improve the sharpness of the roll-off of their designed filters. In continuation of the discussion, the Bridged-T network served as an effective high-pass filter by eliminating even coefficients in the S11 transfer function through the use of identical L1 and L2. Validation through S-parameter simulation using Advanced Design System (ADS) confirms sharp roll-off ratios of -30 dB/GHz for S11 and -32 dB/GHz for S21. This work contributes valuable insights into RF circuit applications and microstrip filter design.



\bibliographystyle{elsarticle-num} 
\bibliography{MyBibDatabase}


%\begin{thebibliography}{00}


%\bibitem{}

%\end{thebibliography}

\end{document}